# CURRENT INJECTION AND VOLTAGE INSERTION ATTACKS AGAINST THE VMG-KLJN SECURE KEY EXCHANGER

SHAHRIAR FERDOUS[1], CHRISTIANA CHAMON, LASZLO B. KISH

*Department of Electrical and Computer Engineering, Texas A&M University, TAMUS 3128, College Station, TX 77841-3128, USA*
*ferdous.shahriar@tamu.edu , cschamon@tamu.edu , laszlokish@tamu.edu*

**Abstract:** In this paper, the vulnerability of the Vadai, Mingesz and Gingl (VMG)-Kirchhoff-Law-Johnson-Noise (KLJN) Key Exchanger (Nature, Science Report 5 (2015) 13653) against two active attacks is demonstrated. The security vulnerability arises from the fact that the effective driving impedances are different between the HL and LH cases for the VMG-KLJN scheme; whereas for the ideal KLJN scheme they are same. Two defense schemes are shown against these attacks but each of them can protect against only one of the attack types; but not against the two attacks simultaneously. The theoretical results are confirmed by computer simulations.

1. **Introduction**

Sensitive data must be encrypted and protected against any kind of breach or eavesdropping during secure communications. Information theoretic (unconditional) security [1-7] provides protection and privacy; irrespective of the computational power, measurement accuracy and speed of the eavesdropper (Eve). At the heart of secure communications is the secure key exchange. Robust, unconditional, and hardware-based secure key exchange is offered by Quantum Key Distribution (QKD) [8-43] and its statistical-physical competitor, the Kirchhoff-Law-Johnson-Noise (KLJN) secure key exchanger [3-7, 44-96]. Quantum Key Distribution (QKD) is based on Quantum physics, particularly the no-cloning theorem; as opposed to KLJN scheme, which is based on classical statistical physics, particularly the Fluctuation Dissipation Theorem [3], Gaussian stochastic process [57] and thermal equilibrium [96].

---

[1] Corresponding Author

*1.1. The KLJN key exchanger*

The core of the KLJN system is shown in Figure 1 [7]. It consists of an information channel which is a wire line between the two communicating parties Alice & Bob; two switches and an identical resistor pair, $R_H$ and $R_L$, (where $R_H > R_L$ and $R_H \neq R_L$) at each of the communicators, Alice and Bob [95]. At the start of the *Bit Exchange Period* (BEP), each party (Alice or Bob) can arbitrarily choose from $R_H$ and $R_L$, and then connect the selected resistor to the wire line for the whole BEP. Alice and Bob publicly agree about the interpretation of the bit value (0 or 1) for the HL ($R_H$ at Alice and $R_L$ at Bob), and the LH ($R_L$ at Alice and $R_H$ at Bob) situations, respectively. The other two situations, HH and LL, are unimportant because those results will be discarded [3, 5], see below.

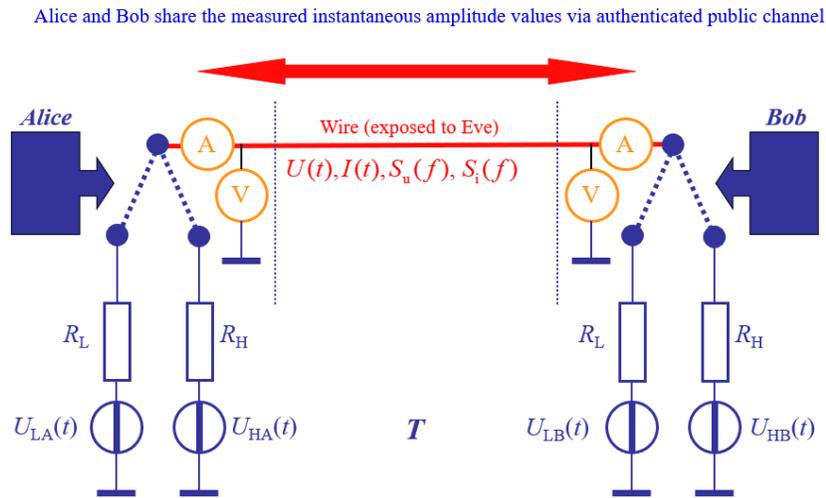

**Fig. 1.** The core of the KLJN secure key exchanger scheme consists of a wire line connection between the two communicating parties Alice & Bob. The voltage $U(t)$, the current $I(t)$, and their spectra $S_u(f)$ and $S_i(f)$, respectively, are measurable by Alice, Bob and Eve. In the private space of Alice and Bob, the voltage generators $U_{LA}(t)$, $U_{HA}(t)$, $U_{LB}(t)$ and $U_{HB}(t)$ represent the independent thermal (Johnson-Nyquist) noises of the resistors, or optionally they are external Gaussian noise generators for higher noise temperature. The homogeneous temperature $T$ in the system guarantees that the LH (Alice $R_L$, Bob $R_H$) and HL (Alice $R_H$, Bob $R_L$) resistor connections provide identical mean-square voltage and current, and the related spectra, in the wire [3,5,95].

The security of the KLJN system is based on the Kirchhoff's Loop Law and the Fluctuation–Dissipation theorem [3, 5-7]. More generally, the unconditional security of the KLJN scheme is derived from the Second Law of Thermodynamics [3], and it requires thermal equilibrium (homogeneous temperature) for the system. The key exchanger utilizes the thermal noise of the resistors that can be emulated by external Gaussian voltage noise generators, too. Since, in thermal equilibrium Alice and Bob operate at equal noise temperatures ($T$), the net power flow between Alice and Bob is zero. Due to the Johnson formula, the HL or LH pairs provide the same mean-square noise voltage spectra and noise current spectra because both the parallel and serial resultant resistances are identical in the two cases.

If, during the BEP, both parties connect the same resistance value HH ($R_H$, $R_H$) or LL ($R_L$, $R_L$), the situation is not secure, see Figure 2, because then Eve knows the connected resistance values at two parties. Thus, the HH & LL bit situations are discarded. The only secure combinations are the HL ($R_H$, $R_L$) and LH ($R_L$, $R_H$) cases [3, 5] because then Eve does not know the locations of the connected resistances.

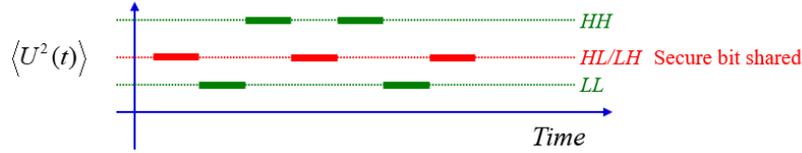

**Fig. 2.** The mean-square value of noise voltage on the wire line versus time during operation [5]. The three different levels (dotted lines) are at HH, HL/LH and LL. Obviously, of the three levels, only the HL/LH level is secure against Eve.

With passive eavesdropping, that is, by measuring the wire voltage and current, Eve can determine the resultant (both parallel and serial) values of the connected resistances by evaluating the noise voltage and current spectra, see below. However, in the HL or LH cases, Eve cannot determine which side has $R_H$ and which side has $R_L$ [3, 7], thus she does not know if the state is HL or LH, which means she does not know if the key bit value is 0 or 1. For Eve this is an information entropy of 1 bit indicating perfect unconditional security.

Specifically, the noise spectra $S_u(f)$ and $S_i(f)$ of the voltage $U(t)$ and current $I(t)$ in the wire, respectively, are given by the Johnson-Nyquist formulas of thermal noise [3, 5]:

$$S_u(f) = 4kTR_p, \tag{1}$$

$$S_i(f) = \frac{4kT}{R_s}, \tag{2}$$

where $k$ is the Boltzmann constant, and $R_p$ and $R_s$ are the parallel and serial resultant values of the connected resistors, respectively. In the HL and LH cases the resultant values are:

$$R_{\text{pLH}} = R_{\text{pHL}} = \frac{R_{\text{L}} R_{\text{H}}}{R_{\text{L}} + R_{\text{H}}} \tag{3}$$

$$R_{\text{sLH}} = R_{\text{sHL}} = R_{\text{L}} + R_{\text{H}}. \tag{4}$$

The quantities that Eve can access with passive measurements satisfy the following equations that, together with Equations (3)-(4), form the pillars of security against passive attacks against the KLJN scheme [95]:

$$U_{\text{LH}} = U_{\text{HL}} \tag{5}$$

$$I_{\text{LH}} = I_{\text{HL}} \tag{6}$$

$$P_{\text{LH}} = P_{\text{HL}} = 0 \quad , \tag{7}$$

where the voltage and current values stand for the effective (RMS) amplitudes in the wire, and $P$ is the mean power flow between Alice and Bob.

### *1.2. The Vadai, Mingesz and Gingl (VMG)-KLJN scheme*

Vadai, Mingesz and Gingl (VMG) introduced a modified scheme in Nature Scientific Reports [47] and proposed that, instead of using identical resistor pairs; we can actually use four *arbitrary* resistors and still maintain perfectly secure communications (see Figure 3). VMG showed that the four arbitrary resistors require (typically) different noise temperatures to guarantee that the voltage and current spectra and the net power flow in the wire are identical for the HL and LH cases. But this implies non-zero mean power flow between Alice and Bob, since the noise temperatures between the resisters are inhomogeneous, so the system is not in thermal equilibrium, anymore. So, it looked like thermal equilibrium was not needed for perfect security, on the contrary of the former understanding [5, 83].

We started investigating these claims in former papers [95, 96] and concluded that these claims are incorrect because there are passive attack types that work against the VMG-KLJN scheme while not against the original KLJN system. In the present paper we are exploring active attacks against the VMG-KLJN scheme, see Sections 2 and 3.

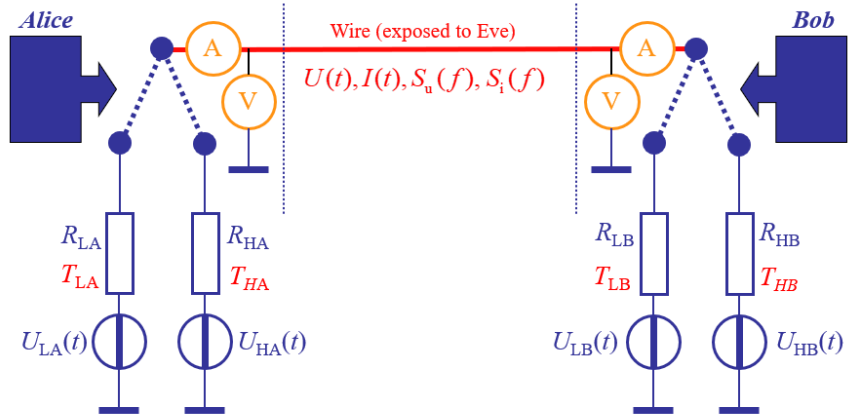

**Fig. 3.** The core of the VMG-KLJN secure key exchanger scheme [47, 95]. The four resistors are different and can be freely chosen (though not totally arbitrarily because of certain unphysical solutions). The voltage generators $U_{LA}(t)$, $U_{HA}(t)$, $U_{LB}(t)$ and $U_{HB}(t)$ represent the thermal noise of the resistors $R_{LA}$, $R_{HA}$, $R_{LB}$ and $R_{HB}$ respectively. $T_{LA}$, $T_{HA}$, $T_{LB}$ and $T_{HB}$ represent the noise temperature of these resistors. The temperature of one of the resistors is freely chosen, and the other 3 temperatures depend on the corresponding resistor values and are given by the VMG Equations (9-11) [47, 95].

For the VMG-KLJN scheme, there are four equations that describe the RMS noise voltages. In search for their solution, they modified Equation (7) by removing the zero power flow conditions [47, 95], as follows:

$$P_{LH} = P_{HL} \,. \tag{8}$$

Then they used Equations (5) and (6) to obtain the necessary mean-square voltages that implies the corresponding temperatures, too [47, 95]:

$$U_{HB}^2 = U_{LA}^2 \frac{R_{LB}(R_{HA}+R_{HB}) - R_{HA}R_{HB} - R_{HB}^2}{R_{LA}^2 + R_{LB}(R_{LA}-R_{HA}) - R_{HA}R_{LA}} = 4kT_{HB}R_{HB}B \tag{9}$$

$$U_{\text{HA}}^2 = U_{\text{LA}}^2 \frac{R_{\text{LB}}(R_{\text{HA}} + R_{\text{HB}}) + R_{\text{HA}}R_{\text{HB}} + R_{\text{HA}}^2}{R_{\text{LA}}^2 + R_{\text{LB}}(R_{\text{LA}} + R_{\text{HB}}) + R_{\text{HB}}R_{\text{LA}}} = 4kT_{\text{HA}}R_{\text{HA}}B \qquad (10)$$

and

$$U_{\text{LB}}^2 = U_{\text{LA}}^2 \frac{R_{\text{LB}}(R_{\text{HA}} - R_{\text{HB}}) - R_{\text{HA}}R_{\text{HB}} + R_{\text{LB}}^2}{R_{\text{LA}}^2 + R_{\text{LA}}(R_{\text{HB}} - R_{\text{HA}}) - R_{\text{HA}}R_{\text{HB}}} = 4kT_{\text{LB}}R_{\text{LB}}B \:. \qquad (11)$$

### *1.3. Former passive attacks against the practical VMG-KLJN scheme*

In [95], we demonstrated two passive attacks against the VMG-KLJN scheme. Our first passive attack [95] was based on the split of crossover frequencies (from white to $1/f^2$ spectrum in the Lorentzian) between the LH and HL situations, when cable capacitance and inductance cannot be ignored. Eve can mathematically calculate the cross-over frequency ($f_{\text{cr}}$) and also measure it and, then compare it between the HL and LH situations [95]. Our second passive attack [95] was based on the split of noise temperature of the line capacitance and inductance. Eve can measure the mean-square noise voltage on the line capacitance ($U_{\text{C}}$) and the mean-square noise current through the line inductance ($I_{\text{L}}$); and use that information to estimate their corresponding noise temperatures, then compare it among the HL and LH situations [95].

Both these attacks can lead to information leak in either the voltage or current or both. And, since these are passive attacks; it will be impossible for Alice/ Bob to discover that Eavesdropping is indeed taking place. In the ideal KLJN scheme, none of these passive attacks exist, because the equivalent parallel or serial resistance between the HL and LH case are identical.

To defend against these two attacks, we proposed two more schemes FCK-2 and FCK-3 [95]. However only one of the attacks can be defended at a time; because mathematically, we cannot have equal parallel and serial resistance at the same time.

Recently, another passive attack, namely the zero-crossing attack has been outlined [96]. Since the VMG-KLJN scheme does not require to be operated under thermal equilibrium, the paper shows that there is a non-zero information leak due to the cross-correlations of the channel's noise voltage and current which is related to zero-crossing time statistics.

*1.3.1. FCK-2: Identical parallel LH and HL resultant resistances: eliminating information leak in the voltage*

When the parallel resultant resistances in the HL and LH cases are identical ($R_{pHL} = R_{pLH}$), then both the corresponding voltage crossover frequencies and the noise voltage temperatures will also be identical [95]. The equation to satisfy this condition is:

$$R_{HB} = \frac{R_{HA} R_{LA} R_{LB}}{R_{HA} R_{LA} - R_{HA} R_{LB} + R_{LA} R_{LB}} \:. \tag{12}$$

*1.3.2. FCK-3: Identical serial LH and HL loop resistances: eliminating information leak in the current*

When the serial resultant resistances in the HL and LH cases are identical ($R_{sHL} = R_{sLH}$), then both the corresponding current crossover frequencies and the noise current temperatures will also be identical [95]. The equation to satisfy this condition is:

$$R_{LB} = R_{LA} + R_{HB} - R_{HA} \:. \tag{13}$$

In the rest of the paper, we will explore active attacks against the VMG-KLJN system.

## 2. Active attack 1: Information leak in the wire voltage due to current injection attack against the VMG-KLJN scheme

Active attack means that Eve injects energy in the system and/or modifies the circuitry. If Eve can do active attacks (either current injection or voltage generator insertion), she can easily crack the VMG-KLJN scheme, due to the imbalance in equivalent HL vs LH resistances, utilizing either the parallel resultant of Alice's and Bob's resistances or the loop (serial) resistance. To prove this claim, we introduce two new active attacks – current injection and voltage insertion (see next section) – against the VMG-KLJN scheme, and we propose defenses against these attacks.

## 2.1. Earlier current injection attacks and defenses against them

Current injection attack, see Figure 4, has been demonstrated in the past in the context for the ideal KLJN scheme [3, 5, 7, 74]. The injected current is distributed unevenly toward Alice and Bob, and the party with the $R_L$ will receive more current than the party with the $R_H$. So, the attack will need *two simultaneous current measurements* in the wire, one toward Alice and another one toward Bob.

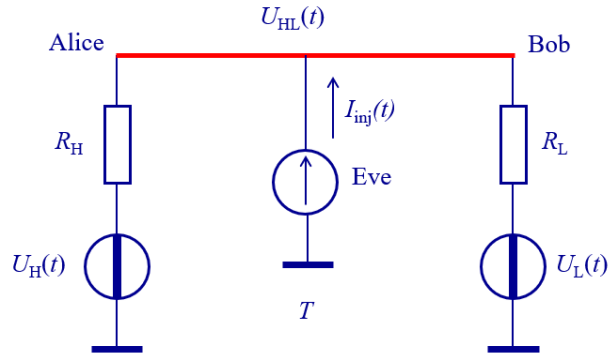

**Fig. 4.** Current injection attack against the ideal KLJN system during the HL case. $U_H(t)$ & $U_L(t)$ are the noise voltages of resistors $R_H$ and $R_L$ respectively; $I_{inj}(t)$ is the injected noise current (of the same noise bandwidth as the resistor noise voltages). Eve measures the current at both directions at the point of the injection.

The concept of the above current injection attack was already introduced in the very first KLJN paper [3]; later the security analysis [7] of the attack was developed. The practical demonstration the attack on a cable model was done in [74]. The proposed defense against the current injection attack has been the monitoring and comparing the channel currents at Alice and Bob [3, 5, 7]. As soon as Alice or Bob observe any discrepancy from the expected channel current, they discard that bit. A more advance protocol was proposed in [7] by running cable simulators by both Alice and Bob; driving it by the measured voltages at the two ends; and comparing the simulated wire currents with the measured ones. An even more advanced defense method was shown in [97], where the same system was used also to prevent active attacks combined with time attacks.

## 2.2. Simplified current injection attack against the VMG-KLJN system

Just like earlier, the current injection attack is carried out by Eve injecting a Gaussian noise current (of same noise bandwidth) into the channel. For attacking the VMG-KLJN scheme, a *single measurement* will be enough: the measurement of the wire voltage at the point of the injection. This attack would not work against the original KLJN scheme because the parallel resistances are identical in the HL and LH cases.

However, VMG uses 4 arbitrary resistors thus the equivalent (parallel) resistance in the HL vs LH case will differ. As an example, assume that for the HL case the total parallel equivalent resistance is higher than for the LH case. Then, in the HL case, the injected current will cause a higher excess mean-square amplitude of the voltage $U_{HL}(t)$ on the wire than that of the voltage $U_{LH}(t)$ in the LH situation.

Similarly to the former current injection attacks (e.g. [3]) the most sensitive information for Eve is obtained not by mean-square amplitude measurements but by the cross-correlation of the injected current and the response in the wire.

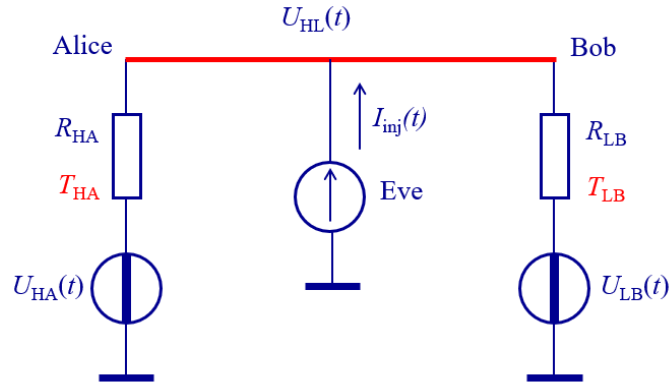

**Fig. 5.** Current injection attack against the VMG-KLJN system during the HL case. $U_{HA}(t)$ & $U_{LB}(t)$ are the noise voltages of resistors $R_{HA}$ and $R_{LB}$ respectively (where $R_{HA}||R_{LB}$ are generally different from $R_{HB}||R_{LA}$); $I_{inj}(t)$ is the injected noise current, and $U_{HL}(t)$ is the total channel noise voltage after the injection.

For the sake of simplicity but without losing generality, we continue to assume that the parallel resistance in the HL-scheme is larger than that of LH-scheme, i.e., $R_{pHL} > R_{pLH}$. The current injection is done via connecting a Gaussian current source $I_{inj}(t)$ (of the same bandwidth as the channel noise) into the channel. During each bit exchange period, Eve is trying to identify HL vs LH situation, by measuring the cross-correlation between the injected current and the voltage on the wire. In the present, HL situation, it is given as:

$$\rho_{\text{HL}} = \langle U_{\text{HL}}(t) * I_{\text{inj}}(t) \rangle, \tag{14}$$

where

$$U_{\text{HL}}(t) = U_{\text{HL0}}(t) + I_{\text{inj}}(t) * R_{\text{pHL}}, \tag{15}$$

and $\langle \ \rangle$ stands for time average.

Here $U_{\text{HL0}}(t)$ is the original wire voltage without the attack. Thus, from Equation (14) it follows:

$$\rho_{\text{HL}} = \langle U_{\text{HL0}}(t) * I_{\text{inj}}(t) + I_{\text{inj}}^2(t) * R_{\text{pHL}} \rangle. \tag{16}$$

Since, $U_{\text{HL0}}(t)$ and $I_{\text{inj}}(t)$ are independent noises, their cross-correlation is zero, thus in the infinite time limit, that term disappears:

$$\rho_{\text{HL}} = \langle I_{\text{inj}}^2(t) \rangle * R_{\text{pHL}}. \tag{17}$$

Similarly, for the LH case:

$$\rho_{\text{LH}} = \langle I_{\text{inj}}^2(t) \rangle * R_{\text{pLH}}, \tag{18}$$

where, for the LH case, $\rho_{\text{LH}}$ is the voltage-current cross-correlation for the LH case.

Using Equations (17) and (18), Eve can calculate the relevant cross-correlations for the HL and LH schemes. Then she compares these results with her measured cross-correlation given by Equation (14), and she can choose between the HL and LH situations.

Alternatively, comparing between the two cross-correlations for a particular key bit, Eve can theoretically obtain:

$$\begin{aligned}\rho_{\mathrm{I}} &= \rho_{\mathrm{iHL}} - \rho_{\mathrm{iLH}} = \left\langle I_{\mathrm{inj}}^2(t)\right\rangle * R_{\mathrm{pHL}} - \left\langle I_{\mathrm{inj}}^2(t)\right\rangle * R_{\mathrm{pLH}} \\ &= \left\langle I_{\mathrm{inj}}^2(t)\right\rangle * \left[R_{\mathrm{pHL}} - R_{\mathrm{pLH}}\right]\end{aligned} \quad , \tag{19}$$

where $\rho_{\mathrm{I}} > 0$ or $\rho_{\mathrm{I}} < 0$ implies that Eve guesses HL or LH for the situation, respectively.

Note, during practical operation, the duration of the time average is not infinite, but it is the BEP. Thus, Eve's signal will be disturbed by the remaining finite-time average of the zero cross-correlation term, and this will contribute to Eve's error probability $e_{\mathrm{E}} = 1 - p_{\mathrm{E}}$, where $p_{\mathrm{E}}$ is the probability of Eve's successful guessing of the shared key bits. See the simulation results in Section 5.

## 3. Active attack 2: Information leak in the wire current due to voltage insertion attack against the VMG-KLJN scheme

For the attack by inserting a Gaussian noise voltage (of same noise bandwidth) into the channel, see Figure 6. This is basically the dual version of the current injection attack. Due to a Reviewer request, here we explain why Eve does not use the electric ground during the voltage attacks and inserts the generator into the loop in a serial fashion. Grounding one of the electrodes of the generator would change the system from a single loop to a double loop, which does not offer any security and the system would not operate.

The inserted voltage will yield different mean-square currents in the HL vs LH case, depending on what is the (serial) loop resistance. Thus, similarly to the current injection, for attacking the VMG-KLJN scheme, a *single measurement* will be enough: the measurement of the wire current. This attack would not work against the original KLJN scheme because the loop resistances are identical in the HL and LH cases.

However, VMG uses 4 arbitrary resistors, thus, in general, the loop resistance in the HL vs LH case will differ. As an example, assume that for the HL case the total serial equivalent resistance is smaller than for the LH case. Then, in the HL case, the inserted voltage will cause a higher excess mean-square amplitude of the current $I_{\mathrm{HL}}(t)$ in the wire than that of the current $I_{\mathrm{LH}}(t)$ in the LH case.

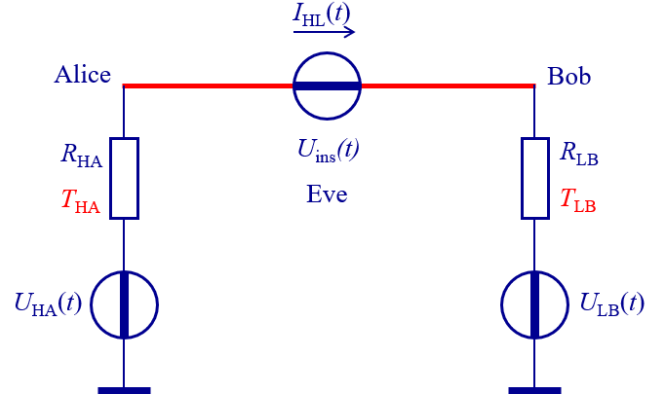

**Fig. 6.** Voltage insertion attack against the VMG-KLJN system during the HL case. $U_{HA}(t)$ and $U_{LB}(t)$ are the noise voltages of resistors $R_{HA}$ and $R_{LB}$ respectively (where $R_{HA}+R_{LB}$ are generally different from $R_{HB}+R_{LA}$); $U_{ins}(t)$ represents the inserted noise voltage, and $I_{HL}(t)$ is the total channel current after the injection.

The most sensitive information for Eve is obtained not by mean-square amplitude measurements but by the cross-correlation of the inserted voltage and the wire current. Here we can progress in the same way as in the current injection attack. For the sake of simplicity, let us continue to assume the effective serial resistor in the HL-scheme is smaller than that of LH-scheme, i.e., $R_{sHL} < R_{sLH}$. From it, we can derive, $\dfrac{1}{R_{sHL}} > \dfrac{1}{R_{sLH}}$.

During each bit exchange period, Eve is trying to identify the HL vs LH situation by measuring the cross-correlation between the inserted voltage and the current in the wire:

$$\rho_{uHL} = \langle I_{HL}(t) * U_{ins}(t) \rangle, \tag{20}$$

where,

$$I_{HL}(t) = I_{HL0}(t) + \frac{U_{ins}(t)}{R_{sHL}}. \tag{21}$$

where $I_{HL0}(t)$ is the wire current without voltage insertion. Thus, from Equation (21) it follows:

$$\rho_{uHL} = \left\langle I_{HL0}(t) * U_{ins}(t) + \frac{U_{ins}^2(t)}{R_{sLH}} \right\rangle, \tag{22}$$

Since $I_{HL0}(t)$ and $U_{ins}(t)$ are independent noises, their cross-correlation is zero thus in the infinite time limit, the first term disappears. What remains is:

$$\rho_{\text{uHL}} = \left\langle \frac{U_{\text{ins}}^2(t)}{R_{\text{sHL}}} \right\rangle, \tag{23}$$

Similarly, for the LH case,

$$\rho_{\text{uLH}} = \left\langle \frac{U_{\text{ins}}^2(t)}{R_{\text{sLH}}} \right\rangle, \tag{24}$$

where, for the LH case, $\rho_{\text{uLH}}$ is the voltage-current cross-correlation for the LH case.

Using Equations (23) and (24), Eve can calculate the relevant cross-correlations for the HL and LH schemes. Then she compares these results with her measured cross-correlation given by Equation (20), and she can choose between the HL and LH situations.

Alternatively, comparing the two cross-correlations for a particular key bit, Eve can theoretically obtain:

$$\rho_{\text{u}} = \rho_{\text{uHL}} - \rho_{\text{uLH}} = \left\langle \frac{U_{\text{ins}}^2(t)}{R_{\text{sHL}}} \right\rangle - \left\langle \frac{U_{\text{ins}}^2(t)}{R_{\text{sLH}}} \right\rangle = \left\langle U_{\text{ins}}^2(t) \right\rangle \left[ \frac{1}{R_{\text{sHL}}} - \frac{1}{R_{\text{sLH}}} \right], \tag{25}$$

where $\rho_{\text{u}} > 0$ or $\rho_{\text{u}} < 0$ imply that Eve must guess HL or LH for the particular situation, respectively.

## 4. Defense methods

We will discuss two types of defense methods to reduce the information leak against each of these active attacks. As we already mentioned, the original KLJN system has zero leak against the above attacks.

- i) Defense by design: Proper selection of the fourth resistor (introduced in this paper).
- ii) Oldest defense: Monitoring and comparing the instantaneous current & voltage amplitudes via an authenticated public communication channel.

### 4.1. Defense by design

Instead of selecting all four resistors arbitrarily, Alice and Bob can choose three resistors arbitrarily and match the impedance of the LH and HL states by properly choosing the fourth resistance [95]. We will demonstrate in Section 5 that this protection eliminates one of the attacks described above but, this approach cannot protect against both attacks simultaneously [95].

#### 4.1.1. Defense against the current injection attack

We can match the LH and HL states by properly choosing the fourth resistor such that, the parallel equivalent resistance becomes equal [95], see Equation (12). When the parallel resultant resistances in the LH and HL cases are identical, the injected current will face equal impedance. Thus, in both the cases the injected current will induce identical voltage on the wire line. This will defend against the current injection attack, as now Eve does not have any way to distinguish which side is HL or LH, by simply measuring the resulting channel voltage.

#### 4.1.2. Defense against the voltage insertion attack

To provide security against the voltage insertion attack, we can match the LH and HL states by properly choosing the fourth resistor such that, the serial equivalent resistance is equal [95], see Equation (13). When the (serial) loop resistances in the LH and HL cases are identical, the inserted voltage will force identical current in the line. This will defend against the voltage insertion attack, as now Eve does not have any way to distinguish which side is HL or LH, by simply measuring the new channel current.

#### 4.1.3. The impossibility to simultaneously fix the parallel and serial resistances:

If the parallel equivalent resistance values are identical for HL and LH, they are different for the serial resistances. Vice versa, if the serial equivalent resistance values are identical for HL and LH, they are different for the parallel resistances [95]. In the VMG-KLJN scheme, therefore it is impossible to fix both the equivalent serial and parallel resistances simultaneously between HL and LH states. Thus at least one of these attacks remains effective.

### 4.2 Defense by current and voltage amplitude monitoring and/or line modeling

The noise generator of Eve, Alice and Bob supply independent stochastic processes [73]. Thus, the injected current or voltage will never cross-correlate with the channel current or voltage.

In order to defend against the current injection (or voltage insertion) attack, Alice and Bob can continuously monitor the instantaneous current (or voltage) at their ends and compare them via the public authenticated data exchange [73-74]. If the monitored instantaneous current and/or voltage do not match the instantaneous value at the other end, then there is an indication of an active attack. Alice and Bob will discard that bit.

In a more advanced fashion [5, 74], after sharing the measured voltage and current data, Alice and Bob can run a cable simulator, where they input the measured voltages at the two end of the cable and simulate the resulting current. If the simulated and measured currents differ, there is a potential for existing active attack. In an even more advanced fashion, this process can be expanded to protect also against the clock attack [97].

## 5. Demonstration of the new active attacks via computer simulations

During the simulations the cable is assumed to be ideal. The generation of Gaussian white noise has been described formerly [86-87]. We have done our demonstrations across three different time steps ($g$) 100, 200, and 500, respectively. For each of the current injection and voltage insertion attack (Table 1 & 3), we simulated 3 different cases of different resistor pair combinations between Alice and Bob. The first case shows an example with ideal KLJN resistor pair, the second case with four arbitrary resistors to show information leak, and the third case with defense methods described in Sections 4.1.1 and 4.1.2. To evaluate Eve's probability $p_E$ of correctly guessing the bit value, we evaluated 2000 BEP.

### 5.1 Simulation results for the simplified current injection attack

The RMS amplitude of the injected current (see Section 2) was 1%, 10% and 20% of the original value in the wire, see Table 1.

Case-A is the classical ideal KLJN situation with HL-Case: $R_{HA}$ = 9kΩ, $R_{LB}$ = 1kΩ, and LH-Case: $R_{LA}$ = 1kΩ, $R_{HB}$ = 9kΩ. As the current injection intensity is increased, $p_E$ still remains around 0.5. To illustrate the accuracy of the simulations, the standard deviation was also calculated. Lastly for a particular $\gamma$, it is seen that the $p_E$ does not change as the injected current is increased. Thus, the attack does not yield any information leak.

Case-B is the simplified current injection attack against the generic VMG-KLJN system. The HL and LH resistor pairs are different and arbitrary, where $R_{HA}$ = 1000Ω, $R_{LB}$ = 160Ω and $R_{LA}$ = 200Ω, $R_{HB}$ = 220Ω; and the corresponding parallel equivalent resistance is 138Ω and 105Ω for the HL and LH case respectively. In the HL and LH cases, the injected current will generate different effective voltages on the channel. We can see at the 1% injected current level that $p_E$ is very close to 0.5; but as we increase the injected current, the $p_E$ value is increasing; indicating that Eve now has a higher probability of identifying the HL or LH state. Thus, as expected, the information leak is more prevalent at higher injection level. Naturally, at higher $\gamma$ values the information leak is larger because of Eve's higher signal-to-noise ratio.

Case-C is the simplified current injection attack against the FCK2-VMG-KLJN system [95]. In this case, three resistors were chosen arbitrarily, where $R_{HA} = 1000\Omega$, $R_{LB} = 160\Omega$, $R_{LA} = 200\Omega$ and we select the fourth resistor ($R_{HB} = 444\Omega$) in such a way that it yields the equivalent parallel resistances to be identical for both the HL and LH states ($R_{pHL} = R_{pLH} = 138\Omega$). Similarly to case-A, the current injection does not create any information leak in the FCK2-VMG-KLJN system [95].

| Case | HL | | LH | | Bit resistance (parallel) | | Injection Factor | $p_E$ $\gamma=100$ | $p_E$ $\gamma=200$ | $p_E$ $\gamma=500$ |
|---|---|---|---|---|---|---|---|---|---|---|
| | $R_{HA}$ k$\Omega$ | $R_{LB}$ k$\Omega$ | $R_{LA}$ k$\Omega$ | $R_{HB}$ k$\Omega$ | $R_{pHL}$ k$\Omega$ | $R_{pLH}$ k$\Omega$ | | | | |
| A | 9 | 1 | 1 | 9 | 0.9 | 0.9 | 1% | 0.504 ± 0.019 | 0.501 ± 0.009 | 0.500 ± 0.015 |
| | | | | | | | 10% | 0.501 ± 0.012 | 0.505 ± 0.008 | 0.503 ± 0.010 |
| | | | | | | | 20% | 0.504 ± 0.008 | 0.501 ± 0.012 | 0.507 ± 0.007 |
| B | 1 | 0.16 | 0.2 | 0.22 | 0.138 | 0.105 | 1% | 0.504 ± 0.011 | 0.504 ± 0.005 | 0.514 ± 0.007 |
| | | | | | | | 10% | 0.525 ± 0.012 | 0.543 ± 0.010 | 0.567 ± 0.010 |
| | | | | | | | 20% | 0.563 ± 0.011 | 0.594 ± 0.009 | 0.635 ± 0.007 |
| C | 1 | 0.16 | 0.2 | 0.444 | 0.138 | 0.138 | 1% | 0.501 ± 0.007 | 0.505 ± 0.009 | 0.503 ± 0.006 |
| | | | | | | | 10% | 0.502 ± 0.010 | 0.505 ± 0.012 | 0.504 ± 0.008 |
| | | | | | | | 20% | 0.509 ± 0.011 | 0.503 ± 0.011 | 0.506 ± 0.012 |

**Table-1.** Demonstration of current injection attack, with different resistor pair combinations, across injection factor (1%, 10%, & 20%) and noise sample size (100, 200 & 500) respectively. $p_E$ is the probability of Eve's successful guessing of the shared key bits. (A) First case is the classical KLJN protocol with identical resistor pairs and identical temperatures of both (HL and LH) resistor pairs, where $R_{HA} = 9000\Omega$, $R_{LB} = 1000\Omega$ and $R_{LA} = 1000\Omega$, $R_{HB} = 100\Omega$; (B) in the second case, the HL and LH resistor pairs are different and arbitrary, where $R_{HA} = 1000\Omega$, $R_{LB} = 160\Omega$ and $R_{LA} = 200\Omega$, $R_{HB} = 220\Omega$; (C) in the third case, three resistors were chosen arbitrarily, where $R_{HA} = 1000\Omega$, $R_{LB} = 160\Omega$, $R_{LA} = 200\Omega$, and we selected the fourth resistor ($R_{HB}$) to be 444.44$\Omega$; in such a way that makes the equivalent parallel resistors identical for both HL and LH case.

Table 2 shows the noise temperatures of the resistors used to demonstrate the current injection attack. For the temperature data [95], noise bandwidth $B = 1000$ Hz and $U_{LA} = 1$V is assumed.

|  | HL | | LH | | Noise Temperature | | | |
| --- | --- | --- | --- | --- | --- | --- | --- | --- |
| Case | $R_{HA}$ k$\Omega$ | $R_{LB}$ k$\Omega$ | $R_{LA}$ k$\Omega$ | $R_{HB}$ k$\Omega$ | $T_{HA}$ (K) | $T_{LB}$ (K) | $T_{LA}$ (K) | $T_{HB}$ (K) |
| A | 9 | 1 | 1 | 9 | $1.81 \times 10^{16}$ | $1.81 \times 10^{16}$ | $1.81 \times 10^{16}$ | $1.81 \times 10^{16}$ |
| B | 1 | 0.16 | 0.2 | 0.22 | $1.70 \times 10^{17}$ | $2.35 \times 10^{16}$ | $9.06 \times 10^{16}$ | $2.09 \times 10^{16}$ |
| C | 1 | 0.16 | 0.2 | 0.444 | $1.31 \times 10^{17}$ | $7.25 \times 10^{16}$ | $9.06 \times 10^{16}$ | $5.82 \times 10^{16}$ |

**Table-2.** Noise temperatures of the resistors to demonstrate the current injection attack. For the temperature data [95], noise bandwidth $B = 1000$ Hz and $U_{LA} = 1$V are assumed. (A) The classical KLJN protocol with identical resistor pairs and identical temperatures of both (HL and LH) resistor pairs with noise temperature $1.81 \times 10^{16}$ K. (B) & (C) The HL and LH resistor pairs are different and arbitrary; thus, their corresponding noise temperatures are also different for each resistor.

## *5.2 Simulation results for the voltage insertion attack*

The RMS amplitude of the inserted voltage (see Section 3) was 1%, 10% and 20% of the original value in the wire, see Table 3.

Case-D is the classical ideal KLJN situation with HL-Case: $R_{HA} = 9$k$\Omega$, $R_{LB} = 1$k$\Omega$; and LH-Case: $R_{LA} = 1$k$\Omega$, $R_{HB} = 9$k$\Omega$. As the voltage insertion intensity is increased, $p_E$ still remains around 0.5. To illustrate the statistical accuracy of the simulations, the standard deviation was also calculated. Lastly for a particular $\gamma$, the $p_E$ does not change as the inserted voltage is increased. Thus, the attack does not yield any information leak.

|  | HL | | LH | | Bit resistance (serial) | | | | | |
| --- | --- | --- | --- | --- | --- | --- | --- | --- | --- | --- |
| Case | $R_{HA}$ k$\Omega$ | $R_{LB}$ k$\Omega$ | $R_{LA}$ k$\Omega$ | $R_{HB}$ k$\Omega$ | $R_{sHL}$ k$\Omega$ | $R_{sLH}$ k$\Omega$ | Injection Factor | $p_E$ $\gamma=100$ | $p_E$ $\gamma=200$ | $p_E$ $\gamma=500$ |
| D | 9 | 1 | 1 | 9 | 10 | 10 | 1% | 0.501± 0.010 | 0.506± 0.009 | 0.503± .011 |
| | | | | | | | 10% | 0.502 ± 0.006 | 0.504 ± 0.014 | 0.504 ± 0.011 |

| | | | | | | | | | |
|---|---|---|---|---|---|---|---|---|---|
| | | | | | | | 20% | 0.503 ± 0.010 | 0.507 ± 0.010 | 0.506 ± 0.010 |
| E | 2 | 2.2 | 0.5 | 2.5 | 4.2 | 3 | 1% | 0.503 ± 0.017 | 0.502 ± 0.011 | 0.512 ± 0.016 |
| | | | | | | | 10% | 0.545 ± 0.009 | 0.566 ± 0.008 | 0.595 ± 0.011 |
| | | | | | | | 20% | 0.582 ± 0.005 | 0.613 ± 0.013 | 0.678 ± 0.006 |
| F | 2 | 1 | 0.5 | 2.5 | 3 | 3 | 1% | 0.500 ± 0.012 | 0.503 ± 0.011 | 0.504 ± 0.010 |
| | | | | | | | 10% | 0.501 ± 0.015 | 0.506 ± 0.010 | 0.500 ± 0.016 |
| | | | | | | | 20% | 0.501 ± 0.013 | 0.501 ± 0.013 | 0.501 ± 0.010 |

**Table-3**. Demonstration of voltage insertion attack, with different resistor pair combinations, across injection factor (1%, 10%, & 20%) and noise sample size (100, 200 & 500) respectively. (D) First case is the classical KLJN protocol with identical resistor pairs and identical temperatures of both (HL and LH) resistor pairs, where $R_{HA} = 9000\Omega$, $R_{LB} = 1000\Omega$ and $R_{LA} = 1000\Omega$, $R_{HB} = 100\Omega$; (E) in the second case, the HL and LH resistor pairs are different and arbitrary, where $R_{HA} = 2000\Omega$, $R_{LB} = 2200\Omega$ and $R_{LA} = 500\Omega$, $R_{HB} = 2500\Omega$; (F) in the third case, three resistors were chosen arbitrarily, where $R_{HA} = 2000\Omega$, $R_{LA} = 500\Omega$, $R_{HB} = 2500\Omega$ and we select the fourth resistor ($R_{LB}$) to be $1000\Omega$; in such a way that makes the equivalent series resistances identical for both HL and LH case.

Case-E is the voltage insertion attack against the generic VMG-KLJN system. The HL and LH resistor pairs are different and arbitrary, where $R_{HA} = 2000\Omega$, $R_{LB} = 2200\Omega$ and $R_{LA} = 500\Omega$, $R_{HB} = 2500\Omega$; and the corresponding serial equivalent resistance is $4200\Omega$ and $3000\Omega$ for the HL and LH case respectively. In the HL and LH cases, the inserted voltage will generate different effective currents on the channel. We can see at the 1% inserted voltage level that $p_E$ is very close to 0.5; but as we increase the inserted voltage, the $p_E$ value is increasing; indicating that Eve now has a higher probability of identifying the HL or LH state. Thus, the information leak is more prevalent at higher insertion level.

Case-F is the voltage insertion attack against the FCK3-VMG-KLJN system [95]. In this case, three resistors were chosen arbitrarily, where $R_{HA} = 2000\Omega$, $R_{LA} = 500\Omega$, $R_{HB} = 2500\Omega$, and we selected the fourth resistor ($R_{LB} = 1000\Omega$) in such a way that it yields the equivalent serial resistances to be identical for both the HL and LH state ($R_{sHL} = R_{sLH} = 3000\Omega$). Similar to case-D, the voltage insertion does not create any information leak in the FCK3-VMG-KLJN system [95].

Table 4 shows the noise temperatures of the resistors used to demonstrate the voltage insertion attack. For the temperature data [95], noise bandwidth $B = 1000$ Hz and $U_{LA} = 1$V is assumed.

|  | HL | | LH | | Noise Temperature | | | |
|---|---|---|---|---|---|---|---|---|
| Case | $R_{HA}$ k$\Omega$ | $R_{LB}$ k$\Omega$ | $R_{LA}$ k$\Omega$ | $R_{HB}$ k$\Omega$ | $T_{HA}$ (K) | $T_{LB}$ (K) | $T_{LA}$ (K) | $T_{HB}$ (K) |
| D | 9 | 1 | 1 | 9 | $1.81 \times 10^{16}$ | $1.81 \times 10^{16}$ | $1.81 \times 10^{16}$ | $1.81 \times 10^{16}$ |
| F | 2 | 2.2 | 0.5 | 2.5 | $2.11 \times 10^{16}$ | $2.31 \times 10^{15}$ | $3.62 \times 10^{16}$ | $2.42 \times 10^{15}$ |
| F | 2 | 1 | 0.5 | 2.5 | $2.72 \times 10^{16}$ | $1.81 \times 10^{16}$ | $3.62 \times 10^{16}$ | $2.17 \times 10^{16}$ |

**Table-4.** Noise temperatures of the resistors used to demonstrate the voltage insertion attack. For the temperature data [95], noise bandwidth $B = 1000$ Hz and $U_{LA} = 1$V is assumed. (D) The classical KLJN protocol with identical resistor pairs and identical temperatures of both (HL and LH) resistor pairs; each resistor is at noise temperature of $1.81 \times 10^{16}$ K. (E) & (F) The HL and LH resistor pairs are different and arbitrary; thus, their corresponding noise temperatures are also different for each resistor.

## *5.3 Simulation results to demonstrate the impossibility to simultaneously defend against current injection and voltage insertion attack:*

It is demonstrated via computer simulations that both $R_s$ and $R_p$ cannot be satisfied at the same time. If we select resistors based on $R_{pHL} = R_{pLH}$, this will protect only against the current injection attack; but the system will be vulnerable against the voltage insertion attack. And, vice versa, if we select resistors based on $R_{sHL} = R_{sLH}$, this will protect against the voltage insertion attack; but the system will be vulnerable against the current injection attack.

Table 5 shows one examples where the VMG-KLJN scheme has identical LH and HL serial resistances and is not immune against the current injection attack. Vice versa, Table 5 also shows another example where the VMG-KLJN scheme has identical LH and HL parallel resistances, and yet, is not immune against the voltage insertion attack. Thus, the four resistors used to protect against current injection attack, does not offer any protection against the voltage insertion attack. Vice versa, the four resistors used to protect against the voltage insertion, remains vulnerable against the current injection attack.

In case-G, we demonstrate current injection attack, with the same resistor pair combination that protected against the voltage insertion attack (obtained during case-F of Table-3).

However, it no longer protects against the current injection attack, where $R_{HA} = 2000\Omega$, $R_{LB} = 1000\Omega$ and $R_{LA} = 500\Omega$, $R_{HB} = 2500\Omega$.

Vice versa, in case-H, we demonstrate voltage insertion attack, with the same resistor pair combination that protected against the current injection attack (obtained during case-C of Table-1). However, it no longer protects against the voltage insertion attack, where $R_{HA} = 1000\Omega$, $R_{LB} = 160\Omega$ and $R_{LA} = 200\Omega$, $R_{HB} = 444\Omega$.

| Case | HL | | LH | | Bit resistance | | Injection Factor | $p_E$ $\gamma=100$ | $p_E$ $\gamma=200$ | $p_E$ $\gamma=500$ |
|---|---|---|---|---|---|---|---|---|---|---|
| | $R_{HA}$ k$\Omega$ | $R_{LB}$ k$\Omega$ | $R_{LA}$ k$\Omega$ | $R_{HB}$ k$\Omega$ | $R_{sHL}= R_{sLH}$ k$\Omega$ | $R_{pHL}= R_{pLH}$ k$\Omega$ | | | | |
| G | 2 | 1 | 0.5 | 2.5 | 3 | N/A | 1% | 0.506 ± 0.008 | 0.505 ± 0.011 | 0.517 ± 0.010 |
| | | | | | | | 10% | 0.533 ± 0.009 | 0.554 ± 0.012 | 0.589 ± 0.011 |
| | | | | | | | 20% | 0.572 ± 0.012 | 0.600 ± 0.010 | 0.661 ± 0.011 |
| H | 1 | 0.16 | 0.2 | 0.444 | N/A | 0.138 | 1% | 0.513 ± 0.014 | 0.506 ± 0.012 | 0.515 ± 0.014 |
| | | | | | | | 10% | 0.546 ± 0.013 | 0.562 ± 0.010 | 0.597 ± 0.009 |
| | | | | | | | 20% | 0.592 ± 0.007 | 0.629 ± 0.009 | 0.689 ± 0.010 |

**Table 5.** (G) Demonstration of current injection attack, with the same resistor pair combination that protected against the voltage insertion attack (obtained during case-F of Table-3), but no longer protects against the current injection attack; where $R_{HA} = 2000\Omega$, $R_{LB} = 1000\Omega$ and $R_{LA} = 500\Omega$, $R_{HB} = 2500\Omega$. (H) Vice versa, demonstration of voltage insertion attack, with the same resistor pair combination that protected against the current injection attack (obtained during case-C of Table-1), but no longer protects against the voltage insertion attack; where $R_{HA} = 1000\Omega$, $R_{LB} = 160\Omega$ and $R_{LA} = 200\Omega$, $R_{HB} = 444\Omega$.

Table 6 shows the noise temperatures of the resistors used in Table 5. For the temperature data [95], noise bandwidth $B = 1000$ Hz and $U_{LA} = 1$V is assumed.

| Case | HL | | LH | | Noise Temperature | | | |
|---|---|---|---|---|---|---|---|---|
| | $R_{HA}$ k$\Omega$ | $R_{LB}$ k$\Omega$ | $R_{LA}$ k$\Omega$ | $R_{HB}$ k$\Omega$ | $T_{HA}$ (K) | $T_{LB}$ (K) | $T_{LA}$ (K) | $T_{HB}$ (K) |
| G | 2 | 1 | 0.5 | 2.5 | $2.72 \times 10^{16}$ | $1.81 \times 10^{16}$ | $3.62 \times 10^{16}$ | $2.17 \times 10^{16}$ |
| H | 1 | 0.16 | 0.2 | 0.444 | $1.31 \times 10^{17}$ | $7.25 \times 10^{16}$ | $9.06 \times 10^{16}$ | $5.82 \times 10^{16}$ |

**Table-6.** Noise temperatures of the resistors used in Table 5. For the temperature data [95], noise bandwidth $B = 1000$ Hz and $U_{LA} = 1$V are assumed. In both cases of (G) & (H), the HL and LH resistor pairs are different and arbitrary; thus, their corresponding noise temperatures are also different for each resistor.

## 6. Conclusion

The VMG-KLJN scheme is vulnerable against the simplified current injection & voltage insertion attacks introduced in this paper, whereas the ideal KLJN scheme is immune to it. By properly choosing the fourth resistor in the loop (the FCK2-VMG-KLJN and the FCK3-VMG-KLJN systems), we can defend against either the current injection or the voltage insertion attack, but not simultaneously.